%% file: generic_rotational_invariant_coding_matrices_for_All_Memoryless_Channels.tex
\documentclass[conference]{IEEEtran}
\ifCLASSINFOpdf
\usepackage[pdftex]{graphicx}
\graphicspath{./figures}
\DeclareGraphicsExtensions{.pdf,.jpeg,.png}
\else
\fi
\usepackage[cmex10]{amsmath}
\usepackage{amssymb}
\usepackage{amsthm}
\usepackage{algorithm}
\usepackage{algorithmic}
\usepackage{array}
\usepackage{url}
\usepackage{cite}
\usepackage{threeparttable}
\usepackage[utf8]{inputenc}  
\usepackage{citesort}
\usepackage[T1]{fontenc} 
\usepackage{mathtools}
\usepackage{amsfonts}
\usepackage{color}
\usepackage{graphicx}
\usepackage{makecell}
\usepackage{etoolbox}
\usepackage{color,soul}
\makeatletter
\patchcmd{\@makecaption}
  {\scshape}
  {}
  {}
  {}
\makeatletter
\patchcmd{\@makecaption}
  {\\}
  {.\ }
  {}
  {}
\makeatother
\def\tablename{Table}

\input{./settings_custom.tex}

\begin{document}
\title{Sparse superposition codes with  rotational invariant coding matrices for memoryless channels}

\author{\IEEEauthorblockN{ YuHao Liu$^{\dagger}$, Teng Fu$^{\dagger}$, Jean Barbier$^{ \diamond }$  and TianQi Hou$^*$\\
$\dagger$ Department of Mathematical Sciences, Tsinghua University, Beijing, China  \\
$\diamond$ International Center for Theoretical Physics, Trieste, Italy    \\
$ * $  Theory Lab, Central Research Institute, 2012 Labs, Huawei Technologies Co., Ltd.\\
} Emails:  \{yh-liu21, fut21\}@mails.tsinghua.edu.cn, jbarbier@ictp.it, thou@connect.ust.hk}
\maketitle
\IEEEpeerreviewmaketitle

\begin{abstract}
 We recently showed in \cite{hou2022sparse} the superiority of certain structured coding matrix ensembles (such as partial row-orthogonal)  for sparse superposition codes when compared with purely random matrices with i.i.d. entries, both information-theoretically and under practical vector approximate message-passing decoding. Here we generalize this result to binary input channels under generalized vector approximate message-passing decoding \cite{schniter2016vector}. We focus on specific binary output channels for concreteness but our analysis based on the replica symmetric method from statistical physics applies to any memoryless channel. We confirm that the ``spectral criterion'' introduced in \cite{hou2022sparse}, a coding-matrix design principle which allows the code to be capacity-achieving in the ``large section size'' asymptotic limit, extends to generic memoryless channels. Moreover, we also show that the vanishing error floor property \cite{biyik2017generalized} of this coding scheme is universal for arbitrary spectrum of the coding matrix. 
\end{abstract}
\section{Introduction}
Since their introduction \cite{barron2010toward} and the proof that they attain the capacity of the additive white Gaussian noise (AWGN) channel \cite{barron2012high,barbier2016proof,rush2017capacity}, sparse superposition (SS) codes have become an active research field given their practical potential \cite{feng2021unifying}, in particular under approximate message-passing (AMP) decoding \cite{barbier2017approximate,rush2017capacity}. But the application range of SS codes has been strongly expanded once it was realized that their desirable properties are universally true for generic memoryless channels when a proper generalization of AMP is employed as a decoder \cite{barbier2016threshold,biyik2017generalized}. Even more recently, following similar studies in the related area of compressive sensing \cite{tulino2013support,ma2014turbo,barbier2018mutual,ma2021analysis}, we have initiated the analysis of SS codes with more generic coding matrices than with i.i.d. entries as is usually the case \cite{hou2022sparse}, but for the AWGN channel only. In the present contribution we go much beyond by extending these latter results to generic memoryless channels. Our main result comes in the form of a simple criterion for the ``optimal'' design among a large class of rotationally-invariant coding matrices, yielding a code which is capacity achieving. Moreover we introduce and analyse the performance of a decoder for SS codes based on the generalized vector approximate message-passing algorithm (GVAMP) \cite{schniter2016vector}. We also show that when decoding is successful it is (asymptotically, but also empirically) perfect for binary input channels: \emph{there is no error-floor}, a very desirable property for any coding scheme. To be concrete we focus on three paradigmatic memoryless
channels: the binary erasure (BEC) and symmetric  (BSC) channels, and the (non-symmetric) Z channel (ZC). But our theory applies to more generic memoryless channels.

Like in \cite{hou2022sparse} our non-rigorous analysis is based on the study of the potential function derived from the replica method \cite{mezard2009information} and its connection to the fixed point(s) of the state evolution (SE) recursions tracking AMP-like algorithms \cite{bayati2011dynamics,javanmard2013state,bayati2011dynamics,javanmard2013state,pandit2020inference,takahashi2022macroscopic}. Nevertheless, a multitude of rigorous studies \cite{rush2017capacity,barbier2020mutual,reeves2016replica,barbier2016threshold,barbier2019optimal,barbier2018mutual} point towards the fact that our predictions should be exact in a proper asymptotic limit. Moreover we empirically confirm through careful numerics that our replica-based theory accurately predicts GVAMP’s performance, i.e., its mean-square error (MSE) after convergence. Therefore our results must be considered as numerically-verified conjectures based on by-now well established techniques from statistical physics.

In SS codes the \emph{message} $\bx \!=\! [\bx_1, \dots, \bx_L]$ is a vector made of $L$ $B$-dimensional \emph{sections}. Each section $\bx_l$, $l\!\in\!\{1,\ldots,L\}$, possesses a single non-zero component equal to $1$ whose position encodes the symbol to transmit. $B$ is the \emph{section size} (or alphabet size) and we set $N\defeq LB$. We consider random codes generated by a \emph{coding matrix} $\bA\!\in\! \mathbb{R}^{M \times N}$ drawn from a rotational invariant ensemble, i.e., when considering its singular value decomposition $\bA = \mathbf{U} \sqrt{\mathbf{S}} \mathbf{V}^\intercal $, the orthogonal bases of singular vectors  $\mathbf{U}$ and $\mathbf{V}$  are sampled uniformly in the orthogonal group $\mathcal{O}(M)$  and $\mathcal{O}(N)$, respectively. The diagonal matrix $ \mathbf{S}$  contains the square of $\mathbf{A}$'s singular values $(S_{i})_{i \leq N}$ on its main diagonal, whose empirical distribution  $N^{-1} \sum_{i \leq N} \delta_{S_{i}} $ weakly converges to a well-defined compactly supported probability density function as $ N, M \to \infty$ (not necessarily proportionally). We denote $\mathbf{A}$'s aspect ratio $\alpha=M/N $ and $ \rho=(1-\alpha) \delta_{0}+\alpha \rho_{\text {supp }} $ the spectral density of $ B^{-1} \mathbf{A}^{\intercal} \mathbf{A} $ as $L\to+\infty$. The cardinality of the code is $B^L$. Hence, the (design) rate is $R=L\log_2(B)/M=\log_2(B)/(\alpha B)$ and thus the code is fully specified by $(M, R, B)$. For a message $\bx$ as before, the \emph{codeword} is $\bA\bx\!\in\! \mathbb{R}^{M}$. We enforce the power constraint $\|\bA\bx\|_2^2/M=1+o_L(1)$ by tuning $\mathbf{A}$'s spectrum so that $ \int d\lambda \lambda \rho_{\text{supp}}(\lambda)= 1$ in the large $L$ limit. The channel $P_{\text {out }}$ outputs the noisy codeword $\by=(y_\mu)_{\mu\le M}$. For the \emph{memoryless} channels we focus on, $P_{\text {out }}(y_{\mu} \mid[\mathbf{A} \mathbf{x}]_{\mu})$ is expressed as
\begin{itemize}
	
	\item BEC: $(1\!-\!\epsilon)\delta(y_\mu\!-\!{\rm sign}([\bA\bx]_\mu))\! +\! \epsilon\delta(y_\mu)$,	
	\item BSC: $(1\!-\!\epsilon)\delta(y_\mu\!-\!{\rm sign}([\bA\bx]_\mu)) \!+\! \epsilon\delta(y_\mu\!+\!{\rm sign}([\bA\bx]_\mu))$,
	\item ZC: $\delta({\rm sign}([\bA\bx]_\mu)\!+\!1)(\epsilon\delta(y_\mu\!-\!1)\!+\!(1\!-\!\epsilon)\delta(y_\mu\!+\!1))\!+\!\delta({\rm sign}([\bA\bx]_\mu)\!-\!1)\delta(y_\mu\!-\!1)$,
\end{itemize}
where $\epsilon$ represents the error probability. The performance measure we are going to analyse is the MSE per section $L^{-1} \mathbb{E}\|\bx - \hat \bx(\by,\bA)\|_2^2$ where $ \hat \bx(\by,\bA)$ will be either the minimum mean-square error (MMSE) or GVAMP estimator.

\section{GVAMP-based decoder for SS codes}
The GVAMP we propose aims at computing the MMSE estimator $\mathbb{E}[\bx\mid \by, \bA]$ given by the mean of the posterior 
 \begin{equation*}
     P(\mathbf{x} \mid  \mathbf{y}, \mathbf{A} ) = \frac{1}{\mathcal{Z}(\mathbf{y}, \mathbf{A})} \prod_{\mu\le M} P_{\text {out}}(y_{\mu} \mid[\mathbf{A} \mathbf{x}]_{\mu}) \prod_{l\le L} P_0(\mathbf{x}_l), 
 \end{equation*}
where $\mathcal{Z}(\mathbf{y}, \mathbf{A})$ is a  normalization. The hard constraints for the sections of the message are enforced by the prior distribution $P_0(\bx_l)=B^{-1}\sum_{i\in l} \delta_{x_i,1}\prod_{j\in l, j\neq i} \delta_{x_j,0}$, where $\{i\!\in\!l\}$ are the $B$ scalar components indices of the section $l$. GVAMP was originally derived for generalized linear estimation \cite{schniter2016vector}. The present generalization to the vectorial setting of SS codes is in the same spirit as the one of AMP for SS codes found in \cite{barbier2014replica}: only the input non-linear step differs from the canonical GVAMP, where the so-called denoiser $ \mathbf{g}_{1}( \mathbf{r},\gamma)$ (which takes into account the prior $P_0$) acts now \emph{section-wise} instead of component-wise. Other than this, the decoder is the standard GVAMP. In full generality it is  $\mathbf{g}_{x1} ( \mathbf{r},\gamma   )\!\defeq\!\mathbb{E}[\mathbf{X} \mid \mathbf{R}\!=\!\mathbf{r}]$ for the random variable $\mathbf{R}\!=\!\mathbf{X}\!+\! \sqrt{\gamma}\,{\mathbf{Z}}$ with $\mathbf{X}\!\sim\!P_0^{\otimes L}$ and ${\mathbf{Z}}\!\sim\!{\cal N}(0,  \tbf{I}_N)$. When plugging $P_0$ in $ \mathbf{g}_{x1}(\mathbf{r},  \gamma ) $ it yields the component-wise expression of the denoiser and its variance:
\begin{align*}
	\begin{cases}
	[ \mathbf{g}_{x1}(\mathbf{r},  \gamma )]_i& := \frac{\exp(r_i/ \gamma)}{\sum_{j\in l_i}\exp(r_j/ \gamma)}, \nonumber\\
	 [ \mathbf{g}_{x1}'(\mathbf{r},  \gamma )]_i  &:=\gamma^{-1} [ \mathbf{g}_{x1}(\mathbf{r},  \gamma )]_{i}( 1- [  \mathbf{g}_{x1}(\mathbf{r},  \gamma )]_{i} ),
	\end{cases}
\end{align*}
where $[\mathbf{g}_{x1}^{\prime}(\mathbf{r},  \gamma )]_i:=[\nabla_\mathbf{r} \mathbf{g} _{x1}(\mathbf{r}, \gamma)]_i$, $l_i$ is the section to which belongs the $i^{\text{th}}$ scalar component. For the auxiliary variable $ \mathbf{z}=\mathbf{A} \mathbf{x}$, in contrast with $\mathbf{g}_{x1}$ that only depends on $P_0$, $\mathbf{g}_{\text{1z}}$ and $\mathbf{g}^{\prime}_{\text{1z}}$ depend on the communication channel model and act component-wise. Their expressions are
\begin{algorithm}[t]
\newcommand{\Kit}{K}
\newcommand{\kp}{k\!+\!}
\caption{GVAMP-based decoder for SS codes}
\begin{algorithmic}[1]  \label{alg:gvamp_slm}
\REQUIRE{$\#$ iterates $K$, coding matrix $\mathbf{A}$, noisy codeword $\mathbf{y}$}
\STATE{Initialize $\mathbf{r}_{1,0}$, $\mathbf{p}_{1,0}$, $\gamma_{1,0}>0$, $\tau_{1,0}>0$.}
\FOR{$k=0,1,\dots,\Kit$ (or until convergence)}
    \STATE{// Denoising $\mathbf{x}$}
    \STATE{$\hat{\mathbf{x}}_{1, k}=\mathbf{g}_{x 1}\left(\mathbf{r}_{1, k}, \gamma_{1, k}\right), \quad \alpha_{1, k}=\left\langle\mathbf{g}_{x 1}^{\prime}\left(\mathbf{r}_{1, k}, \gamma_{1, k}\right)\right\rangle$}
    \STATE{$\mathbf{r}_{2, k}=\left(\hat{\mathbf{x}}_{1, k}-\alpha_{1, k} \mathbf{r}_{1, k}\right) /\left(1-\alpha_{1, k}\right)$}
    \STATE{$\gamma_{2, k}=\gamma_{1, k}\left(1-\alpha_{1, k}\right) / \alpha_{1, k}$}
    
    \STATE{// Denoising $\mathbf{z}$}
    \STATE{$\hat{\mathbf{z}}_{1, k}=\mathbf{g}_{z 1}\left(\mathbf{p}_{1, k}, \tau_{1, k}\right), \quad \beta_{1, k}=\left\langle\mathbf{g}_{z 1}^{\prime}\left(\mathbf{p}_{1, k}, \tau_{1, k}\right)\right\rangle$}
    \STATE{$\mathbf{p}_{2, k}=\left(\hat{\mathbf{z}}_{1, k}-\beta_{1, k} \mathbf{p}_{1, k}\right) /\left(1-\beta_{1, k}\right)$}
    \STATE{$\tau_{2, k}=\tau_{1, k}\left(1-\beta_{1, k}\right) / \beta_{1, k}$}
    
    \STATE{// LMMSE estimation of $\mathbf{x}$}
    \STATE{$\hat{\mathbf{x}}_{2, k}=\mathbf{g}_{x 2}\left(\mathbf{r}_{2, k}, \mathbf{p}_{2, k}, \gamma_{2, k}, \tau_{2, k}\right)$}  \label{line:x2}
    \STATE{ $\alpha_{2, k}=\left\langle\mathbf{g}_{x 2}^{\prime}\left(\mathbf{r}_{2, k}, \mathbf{p}_{2, k}, \gamma_{2, k}, \tau_{2, k}\right)\right\rangle$} \label{line:alpha2}
    \STATE{$\mathbf{r}_{1, k+1}=\left(\hat{\mathbf{x}}_{2, k}-\alpha_{2, k} \mathbf{r}_{2, k}\right) /\left(1-\alpha_{2, k}\right)$}
    \STATE{$\gamma_{1, k+1}=\gamma_{2, k}\left(1-\alpha_{2, k}\right) / \alpha_{2, k}$}
    
    \STATE{// LMMSE estimation of $\mathbf{z}$}
    \STATE{$\hat{\mathbf{z}}_{2, k}=\mathbf{g}_{z 2}\left(\mathbf{r}_{2, k}, \mathbf{p}_{2, k}, \gamma_{2, k}, \tau_{2, k}\right)$}
    \STATE{$\beta_{2, k}= \left\langle\mathbf{g}_{z 2}^{\prime}\left(\mathbf{r}_{2, k}, \mathbf{p}_{2, k}, \gamma_{2, k}, \tau_{2, k}\right)\right\rangle$} \label{line:beta2}
    \STATE{$\mathbf{p}_{1, k+1}=\left(\hat{\mathbf{z}}_{2, k}-\beta_{2, k} \mathbf{p}_{2, k}\right) /\left(1-\beta_{2, k}\right)$}
    \STATE{$\tau_{1, k+1}=\tau_{2, k}\left(1-\beta_{2, k}\right) / \beta_{2, k}$} \label{line:tau1}
\ENDFOR
\STATE{Return $\hat{\mathbf{x}}=\hat{\mathbf{x}}_{1, K}$.}
\end{algorithmic}
\end{algorithm}
\begin{equation*}
   \mathbf{g}_{z 1}(\mathbf{p},  \tau )  := \mathbb{E}_{p}\, \textbf{z}, \qquad  \mathbf{g}_{z 1}'(\mathbf{p},  \tau )   :={\rm Cov}_{p}\, \textbf{z} , 
\end{equation*}
where the expectation and covariance matrix are taken with respect to $p(\mathbf{z} \mid\by) \propto P_{\text{out}} (\mathbf{y} \mid \mathbf{z}) \mathcal{N}(\mathbf{z}; \mathbf{p},\mathbf{I} / \tau) $ (where $\mathcal{N}(\mathbf{z};\mathbf{a},\mathbf{b})$ is the probability density function of the normal distribution with mean $\mathbf{a}$ and covariance $\mathbf{b}$). The LMMSE estimators $ \mathbf{g}_{x 2} $ and $ \mathbf{g}_{z 2} $ are related to the following pseudo linear model: $\bar{\by}=\bar{\bA} \bar{\bx}+\bar{\bw}$ where $\bar{\by} := \boldsymbol{0}, \bar{\bA} := \mat{\bA-\mathbf{I}_{\text{M} }}, \bar{\bx} :=\smallmat{\bx\\ \bz} $, and $\bar{\mathbf{w}} \sim \mathcal{N}(\mathbf{0}, \mathbf{I}_{\text{M}} / \gamma_{e})$ with prior $ \bar{\mathbf{x}} \sim \mathcal{N}(\smallmat{\mathbf{r}_{2 k} \\ \mathbf{p}_{2 k}},[\begin{smallmatrix}
\mathbf{I}_{\text{N} } / \gamma_{2 k} & \boldsymbol{0} \\
\boldsymbol{0} & \mathbf{I}_{\text{M} } / \tau_{2 k}
\end{smallmatrix}])$. The LMMSE estimate is $\int d\bar{\bx}\, \bar{\bx}\,  p(\bar{\bx} \mid \bar{\by})$, where $p(\bar{\bx} \mid \bar{\by}) \propto p(\bar{\by}\mid \bar{\bx}) p(\bar{\bx})$. Then in the limit $\gamma_{e}  \to \infty $ strictly enforcing $ \mathbf{z}=\mathbf{A} \mathbf{x} $, the LMMSE estimate and its variances read \cite{schniter2016vector}
\begin{equation*}
	\begin{cases}
    \mathbf{g}_{x 2}(\mathbf{r}_{2, k}, \mathbf{p}_{2, k}, \gamma_{2, k}, \tau_{2, k}):=\boldsymbol{K}^{-1} (\tau_{2,k} \mathbf{A}^\intercal \mathbf{p}_{2,k} + \gamma_{2,k} \mathbf{r}_{2,k}),\\
    \mathbf{g}_{z 2}(\mathbf{r}_{2, k}, \mathbf{p}_{2, k}, \gamma_{2, k}, \tau_{2, k}):=\mathbf{A} \mathbf{g}_{x 2}(\mathbf{r}_{2, k}, \mathbf{p}_{2, k}, \gamma_{2,k}, \tau_{2,k}),
\end{cases}
\end{equation*}
where $\boldsymbol{K} := \tau_{2,k} \mathbf{A}^\intercal \mathbf{A} + \gamma_{2,k} \mathbf{I}$. Moreover we have
\begin{equation*}
\begin{cases}
    \mathbf{g}_{x 2}^{\prime}\left(\mathbf{r}_{2, k}, \mathbf{p}_{2, k}, \gamma_{2, k}, \tau_{2, k}\right) = \gamma_{2,k} \boldsymbol{K}^{-1}\\
    \mathbf{g}_{z 2}^{\prime}\left(\mathbf{r}_{2, k}, \mathbf{p}_{2, k}, \gamma_{2, k}, \tau_{2, k}\right) =  \tau_{2,k} \mathbf{A} \boldsymbol{K}^{-1} \mathbf{A}^\intercal.
    \end{cases}
\end{equation*}
where the prime $'$ means derivative w.r.t. the first argument, and
$\langle \boldsymbol{M} \rangle = k^{-1} \operatorname{Tr} \boldsymbol{M}$ for a matrix  $\boldsymbol{M} \in \mathbb{R}^{k \times k}$, or $\langle \boldsymbol{m} \rangle = k^{-1} \sum_{i\le k} m_i$ for $\boldsymbol{m}\in \mathbb{R}^k$.

\section{Asymptotic analysis by the replica method}

The performance of SS codes in the $L \to  \infty$ limit will be analyzed using the non-rigorous (yet conjectured exact) replica method -- which, again, has been proved to be correct in many inference problems \cite{barbier2020mutual,reeves2016replica,barbier2019optimal,barbier2018mutual,lelarge2019fundamental,dia2016mutual,gerbelot2020asymptotic} -- in order to obtain both the minimum mean-square error and GVAMP's fixed point performance. Note that we do not aim at tracking its per-iterate performance, which would instead require to use the rather involved state evolution analyses of \cite{takahashi2022macroscopic,pandit2020inference}. Actually, making SE rigorous for SS codes (a goal beyond the scope of the present paper) requires special care, see \cite{rush2017capacity}. So even if we were using the previous references to track GVAMP by state evolution, it would not be rigorous (even if probably correctly tracking the decoder for any practical purpose) and so we would not gain much compared to our replica approach. Our choice of using the replica method only is thus that $i)$ it allows to access both performance measures (algorithmic and information-theoretic), and $ii)$ despite their apparent technicality, our replica equations remain simpler than GVAMP's state evolution \cite{takahashi2022macroscopic,pandit2020inference}, the reason being that our equations focus on the fixed point rather than on dynamics.

\subsection{Replica potential function}
The goal of the replica method is to compute the so-called \emph{free entropy} (i.e., log-partition function) $\Phi := \mathbb{E}_{\by,\mathbf{A}}\ln \mathcal{Z}(\mathbf{y},\mathbf{A})$ using the ``replica trick'' $\mathbb{E}\ln \mathcal{Z}=\lim_{n\to 0}\partial_n\ln\mathbb{E} \mathcal{Z}^n$. For any fixed $B\ge 2$ we adapt the results of \cite{takahashi2022macroscopic,shinzato2008perceptron,maillard2020phase} for the standard GLM with generic rotational invariant matrices to SS codes, namely, we make the necessary changes required to go from a scalar setting ($B=1$) to the section-wise setting of SS codes ($B\ge 2$); the complete derivation will be reported in a longer version. The resulting variational formula reads
\begin{align}
    &\Phi =\sup_{ q_{x} \in[0, \frac{1}{B}],q_z \ge 0}\inf_{\hat{q}_x \geq 0 ,\hat{q}_z  \in[0, 1]} \Phi_{\rm RS}(q_x,q_z,\hat q_x, \hat q_z) \label{freeEnt},\\
    &\Phi_{\rm RS}(q_x,q_z,\hat q_x, \hat q_z):= I_0(q_{x},\hat q_x) + \alpha I_{\text{out}}(q_{z},\hat q_z) + I_{\text{int}}(q_{x},q_{z}),\nonumber
\end{align} 
where the functions constructing the \emph{replica potential} $\Phi_{\rm RS}$ are
\begin{align*}
\begin{cases}
I_0(q_x,\hat q_x):=  \mathbb{E}_{\boldsymbol{\xi},\bS}  \ln \mathcal{Z}_0(\hat q_x, \bS,\bxi)-\frac{B}{2}q_x \hat{q}_x , \\
    I_{\text{int}}(q_x,q_z):= B \mathcal{F}(1-Bq_x,q_z)+ \frac{1}{2} \alpha B q_z,\\ I_{\text{out}}(q_z,\hat q_z):= B \mathbb{E}_\xi\int   d y  \mathcal{Z}_{\text {out }}(y;\hat{q}_z \xi, 1-\hat{q}_{z} ) \\ 
    \qquad \qquad\qquad \times\ln  \mathcal{Z}_{\text {out}}(y;\hat{q}_z \xi, 1-\hat{q}_{z}  ) -\frac{B}{2} q_z \hat{q}_z.
\end{cases}    
\end{align*}
with $\xi\sim \mathcal{N}(0,1)$, $\boldsymbol{\xi} \sim \mathcal{N}(0,\mathbf{I}_B)$ and $\mathbb{R}^B\ni \bS,\bs\sim P_0$ all independently. The auxiliary functions are 
\begin{align*}
    \begin{cases}
    \mathcal{Z}_0(\hat q_x, \bS,\bxi) :=\mathbb{E}_{\bs}\exp\big(-\frac{\hat q_x}{2} \|\mathbf{s}\|^2_2 +\sqrt{\hat q_x}\mathbf{s}^\intercal (\sqrt{\hat q_x}\bS +  \bxi) \big), \\
    \mathcal{Z}_{\text {out }}(y;\omega, v ):=\int dz P_{\text{out}}(y\mid z)\mathcal{N}(z; \omega,v),
    \end{cases}
\end{align*}
and $\mathcal{F}(x,y)$ is the \emph{rectangular spherical integral} used, e.g., in \cite{maillard2020construction,ccakmak2020dynamical} where it is expressed as follows:
\begin{align}
        2\mathcal{F}(x,y) &:=\inf_{\Lambda_{x}, \Lambda_{y}  \geq 0 } \big\{ (1-\alpha)\ln \Lambda _y-\mathbb{E} \ln \left( \Lambda _x\Lambda _y+\lambda \right)  \nonumber\\
         &  +\Lambda _xx+{\alpha \Lambda _yy}
            -\ln x-\alpha\ln y-\alpha -1\big\},\label{rectSph}
\end{align}
where $\lambda\sim \rho$ with $\rho$ the asymptotic spectral density of $ B^{-1}\mathbf{A}^\intercal \mathbf{A}$. For i.i.d. Gaussian ensembles, whose spectrum density is the Marcenko-Pastur (MP) law, $\mathcal{F}_{\text{MP}}(x,y)=- \frac{\alpha}2 xy$; for the row-orthogonal ensemble with spectral density $\rho_{\text{row}}  = (1-\alpha) \delta_0 + \alpha \delta_1$ it is $\mathcal{F}_{\text{row}}(x,y) = -\frac{1}{2}\ln(\frac12(1+\sqrt{1-4\alpha xy}))+\frac12\sqrt{1-4\alpha xy}-\frac{1}{2}$. So we have a decomposition of the potential into a part $I_0$ encoding information about the prior $P_0$, $I_{\text {out}}$ on the channel $P_{\text{out}}$ and $I_{\text{int}}$ on the coding ensemble through $\rho$.

\subsection{Stationary equations of the replica potential}

Assuming that the various extrema of the above variational problems are attained inside the optimization domains, the coupled stationary equations obtained by setting $\nabla\Phi_{\rm RS}=\boldsymbol{0}$ read (again $\mathbb{R}^B\ni\bs\sim P_0$, $\boldsymbol{\xi} \sim \mathcal{N}(0,\mathbf{I}_B)$ and $\xi \sim \mathcal{N}(0,1)$)
\begin{align*}
       \begin{cases}
        q_{x}=B^{-1}\mathbb{E}_{\bs,\bxi}\|\mathbb{E}[\bs \mid \by= \sqrt{\hat q_x} \bs +\bxi]\|^2_2,\\ 
        q_{z}=\mathbb{E}_{\xi} \int d y \mathcal{Z}_{\text {out }}(y; \sqrt{\hat{q}_{z}}\xi, 1-\hat{q}_{z})\\
        \qquad\qquad\qquad\times\, |\partial_{\omega} \ln \mathcal{Z}_{\text{out}}(y ; \omega, 1-\hat{q}_{z})|_{\omega=\sqrt{\hat{q}_{z}}\xi}|^{2},  \\
        \hat{q}_{x}=2 \partial q_{x} \mathcal{F}(1-B q_{x}, q_{z}), \\ 
        \hat{q}_{z}=1+2 \alpha^{-1}\partial q_{z} \mathcal{F}(1-B q_{x}, q_{z}).
       \end{cases}
\end{align*}
The above stationary conditions of the replica potential will be our main tool of analysis (we call the first the $q_x$-stationary equation, etc.). Indeed, a powerful feature of the variational formula \eqref{freeEnt} is that the associated stationary conditions can characterize both the MMSE and the MSE attained by the GVAMP algorithm after convergence in the limit $L\to+\infty$ \cite{takahashi2022macroscopic,pandit2020inference} as we describe in the next section. 

In particular, the ``overlap'' $q_x$ physically corresponds to the inner product $\lim_{L\to+\infty} N^{-1}\mathbb{E}[\bx^\intercal \hat \bx]$ between the signal $\bx$ and $\hat \bx$ that can be either the MMSE or GVAMP estimator. Therefore, given one solution $(q_x,q_z,\hat q_x,\hat q_z)$ of the stationary equations (which, as we will see, can characterize both the MMSE or GVAMP estimators), the replica prediction for the corresponding asymptotic MSE per section is $1-B q_x$. Simplifying the $q_x$-stationary equation we get an equivalent expression for this MSE which this time depends on $\hat q_x$ and which is more practical/stable when $q_x$ becomes small:
\begin{align}
   &E(\hat q_x):= \mathbb{E}_{\bxi }\big[(f_{1}(\hat{q}_x,\boldsymbol{\xi})-1)^2 + (B-1)f_{2}(\hat{q}_x,\boldsymbol{\xi})^2\big]\label{MSE},\\
   &\begin{cases*}
     f_{1}(x,\boldsymbol{\xi}) := (1+e^{-x} \sum_{i=2}^{B} e^{\sqrt{x} ({\xi}_i-{\xi}_1)})^{-1}, \\ f_{2}(x,\boldsymbol{\xi}) :=(1+e^{x+\sqrt{x}(\xi_1-\xi_2)}+e^{\sum_{i=3}^{B} \sqrt{x}(\xi_i-\xi_2)})^{-1}.      
    \end{cases*}
\end{align}
For a solution $(q_x,q_z,\hat q_x,\hat q_z)$, $E(\hat q_x)=1-Bq_x$, see \cite{barbier2017approximate}.

\subsection{Analyzing the replica stationary equations}
\begin{figure}[htbp]
	\centering
	 \includegraphics[trim={70pt 83pt 55pt 93pt}, clip, width=0.53\textwidth]{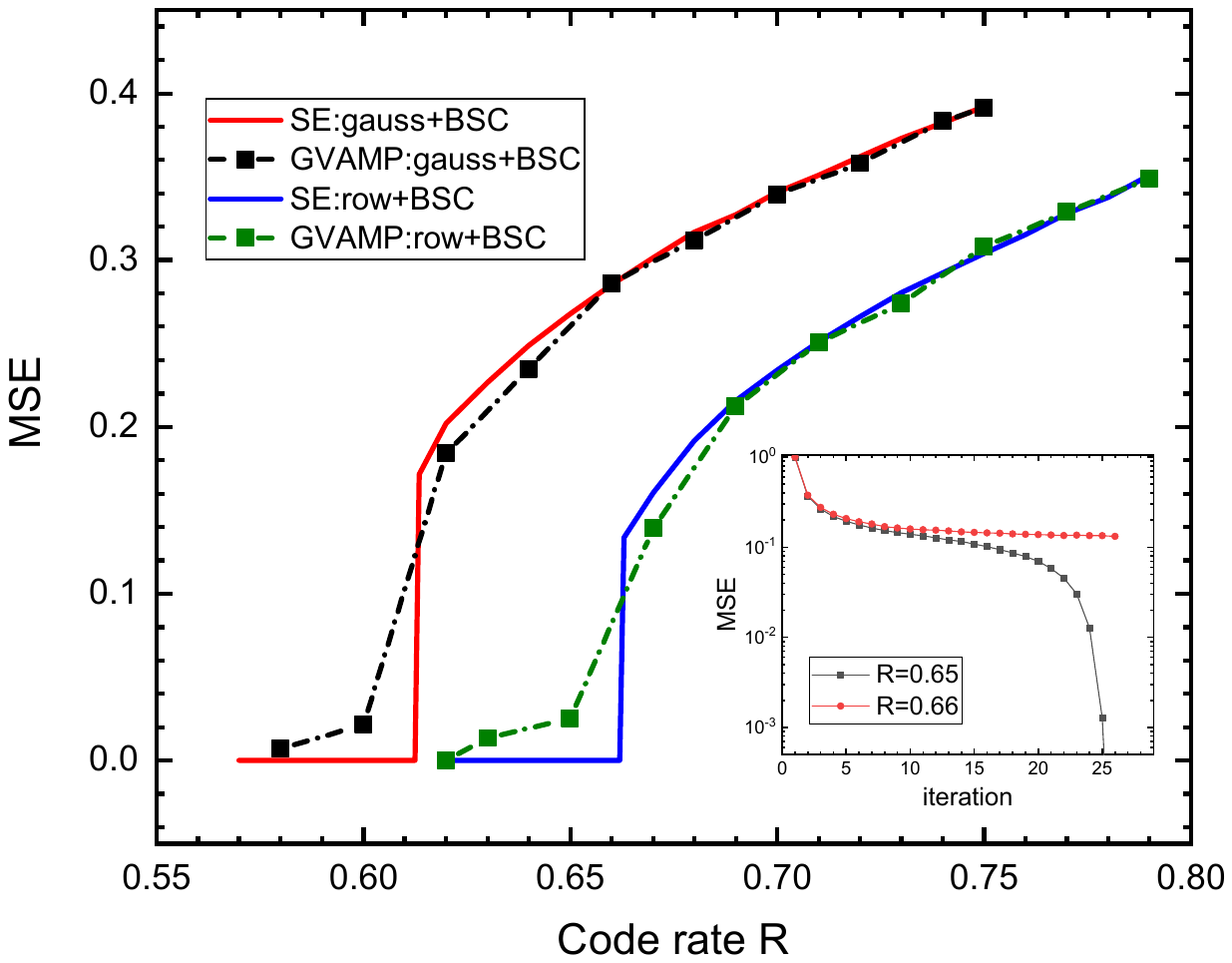}
	 \includegraphics[trim={70pt 53pt 55pt 93pt}, clip, width=0.53\textwidth]{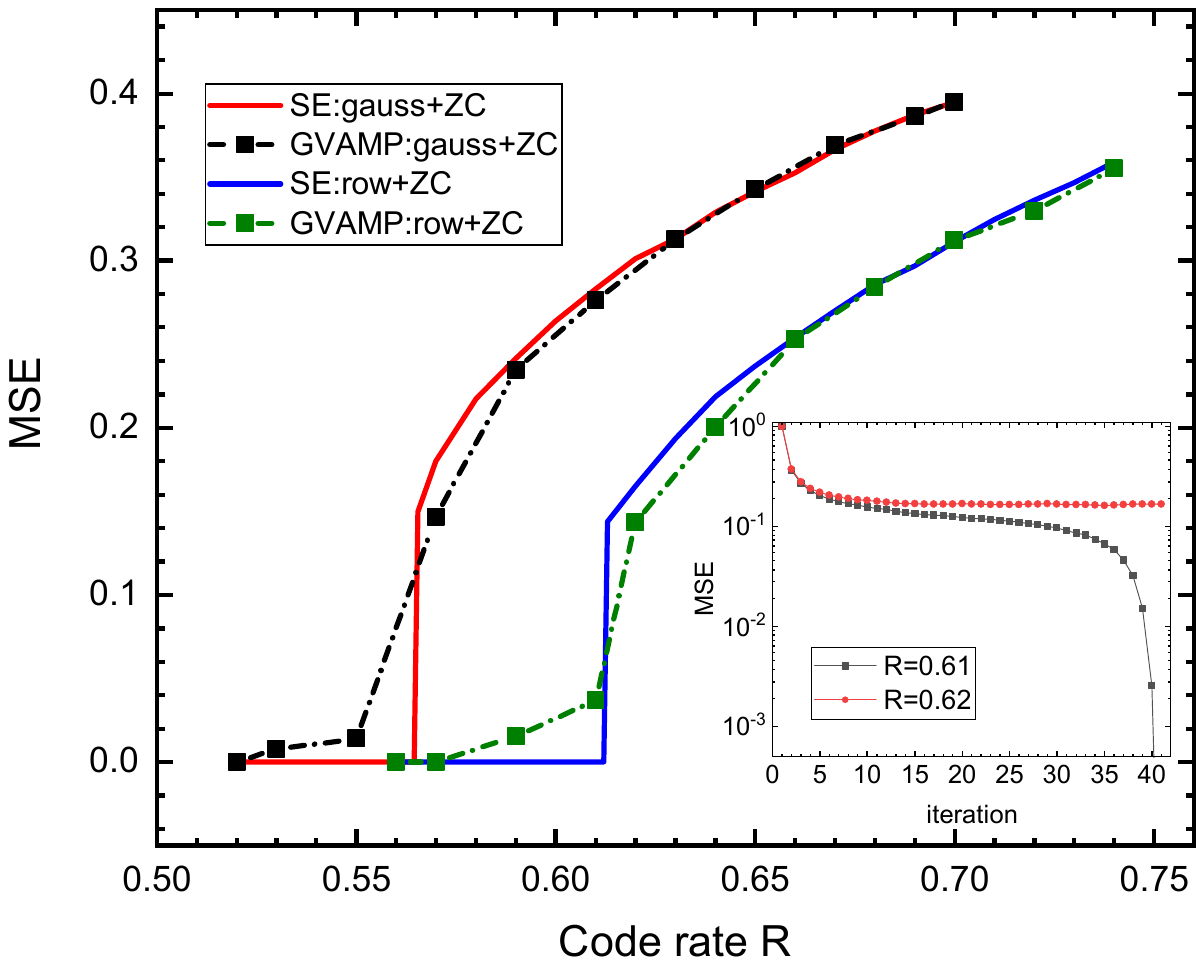}
	 \caption{The solid lines shows the GVAMP asymptotic fixed point MSE predicted from the replica stationary equations. The dashed dot lines are finite size performances of GVAMP over the $\text{BSC}(\epsilon = 0.01)$ and $\text{ZC}(\epsilon = 0.05)$ averaged over 50 instances with $L=2^{14}$ and $B=8$, as a function of the code rate $R$. The stationary equations are solved by Monte Carlo integration with $5 \times 10^6$ samples. Two types of coding matrices are considered: standard coding matrices with i.i.d. Gaussian entries, and partial row-orthogonal ones. It is observed that as predicted by the theory the MSE error floor vanishes when the rate is smaller than their respective algorithmic threshold $R_{\text{GVAMP}}$, showed in table \ref{table:gampForSS}. Clearly GVAMP performs better for row-orthogonal ensembles. The subfigures show the GVAMP iterates on one instance of size $L=2^{14}$ and $B=8$, with row-orthogonal matrices, as a function of the iterations. }\label{fig}
	 \vspace{-10pt}
\end{figure}

The MMSE and GVAMP performances are obtained by iteratively solving the stationary equations starting from two distinct initial conditions: the \emph{informative intialization} is $(q_{{\rm in},x}^{t=0}=B^{-1}, q_{{\rm in},z}^{t=0} >0)$ and gives access to solution $\mathbf{q}_{\rm in}=(q_{{\rm in},x}^{\infty},q_{{\rm in},z}^{\infty},\hat{q}_{{\rm in},x}^{\infty},\hat{q}_{{\rm in},z}^{\infty})$. Because of the aforementioned MSE--$q_x$ connection, algorithmically this means ``initializing on the solution'', i.e., an oracle initialization with MSE $1- B q_{{\rm in},x}^{t=0}=0$. Instead the \emph{un-informative intialization} $(q_{{\rm un},x}^{t=0}=0, q_{{\rm un},z}^{t=0} >0)$ yields the solution $\mathbf{q}_{\rm un}$ which verifies $q^\infty_{{\rm un },x}\le q^\infty_{{\rm in },x}$. It corresponds to a practical initialization without knowledge of the signal. We empirically verified that only these two fixed points exist, independently of how $q_z>0$ is initialized. This holds in more standard settings of SS codes \cite{barbier2017approximate}. With these two solutions in hand, one needs to plug each of them in the replica potential $\Phi_{\rm RS}$ and compare the obtained values; the reason for that step is explained below. 
Denote $\Phi_{\rm in}:=\Phi_{\rm RS}(\mathbf{q}_{\rm in})$ and $\Phi_{\rm un}:=\Phi_{\rm RS}(\mathbf{q}_{\rm un})$ (keep in mind that these are functions of the rate $R$). In the replica theory, the MMSE is extracted from the fixed point with the highest free entropy (the so-called ``thermodynamic equilibrium state'' in physics parlance). So, denoting $\mathbf{q}_{\rm opt}:={\rm argmax}_{\mathbf{q}\in\{\mathbf{q}_{\rm in },\mathbf{q}_{\rm un }\}} \Phi_{\rm RS}(\mathbf{q})$, the replica prediction for the asymptotic MMSE is
$$\lim_{L\to +\infty}L^{-1} \mathbb{E}\|\bx - \mathbb{E}[\bx \mid \by,\bA]\|_2^2= 1-B{q}_{{\rm opt},x}^\infty= E(\hat{q}_{{\rm opt},x}^\infty).$$ 
Instead, GVAMP's MSE is given by plugging in it $\hat{q}_{{\rm un},x}^\infty$:
$$\lim_{L\to +\infty}L^{-1} \mathbb{E}\|\bx - \hat \bx_\text{GVAMP}(\by,\bA)\|_2^2= 1-B{q}_{{\rm un},x}^\infty= E(\hat{q}_{{\rm un},x}^\infty).$$ 
This means that, as pointed in \cite{takahashi2022macroscopic,pandit2020inference}, the fixed point of the state evolution recursions describing GVAMP's MSE (and therefore its MSE for finite but large sizes $L$) can be accessed via the (simpler to implement) above equations. This is confirmed numerically, see Fig.~\ref{fig}.\\

\noindent \textbf{Phase diagram for SS codes} \ \ Depending on the rate $R$ distinct regions exist and the transitions between them define two thresholds that can be extracted from the replica potential: the \emph{GVAMP algorithmic threshold} $R_\text{GVAMP}$ and the \emph{information-theoretic threshold} $R_\text{IT}$:
\begin{align*}
R_\text{GVAMP}&:=\inf\{R\!:\! \Phi_{\rm un}\!<\! \Phi_{\rm in}\},\  R_{\text{IT}}:=\sup\{R\!:\! \Phi_{\rm un}\!<\! \Phi_{\rm in}\}.
\end{align*}
Their analysis is one of our main goal. Equipped with the replica potential and these definitions, we describe the phase diagram (as $R$ increases) using $\mathbf{q}_{\rm in },\mathbf{q}_{\rm un },\Phi_{\rm in},\Phi_{\rm un}$:

\noindent $\bullet$ {\bf Easy phase} $R<R_\text{GVAMP}$:  In this region $\mathbf{q}_{\rm in}=\mathbf{q}_{\rm un}$ and thus GVAMP achieves the MMSE $1-B q_{{\rm un},x}^\infty=E(\hat q^\infty_{{\rm un },x})$ which is ``small''. Decoding is computationally efficient using GVAMP. At a higher rate than $R_\text{GVAMP}$ the fixed points differ and we enter the computationally hard phase. Threshold $R_\text{GVAMP}$ corresponds to the rate where the solid curve(s) jumps discontinuously on Fig.~\ref{fig}.

\noindent $\bullet$ {\bf Hard phase} $R_\text{GVAMP}< R<R_\text{IT}$: In this region $\mathbf{q}_{\rm in }\neq \mathbf{q}_{\rm un }$ and $\Phi_{\rm un}< \Phi_{\rm in}$. GVAMP is sub-optimal, i.e., a statistical-to-computational gap is present. The MMSE equals $1- B q^\infty_{{\rm in },x}=E(\hat q^\infty_{{\rm in },x})$ and is strictly lower then GVAMP's MSE $1- B q^\infty_{{\rm un },x}=E(\hat q^\infty_{{\rm un },x})$. Beyond $R_\text{IT}$ the quality of inference becomes poor using any procedure, efficient or not.

\noindent $\bullet$ {\bf Impossible phase} $R>R_\text{IT}$:   $\mathbf{q}_{\rm in }$ may be equal or not to $\mathbf{q}_{\rm un }$ but the free entropy $\Phi_{\rm un}\ge \Phi_{\rm in}$ and $q_{{\rm un},x}^\infty$ is ``small''. In this case GVAMP is optimal and its MSE $1-Bq_{{\rm un},x}^\infty=E(\hat q_{{\rm un},x}^\infty)$ (which matches the MMSE) is ``large''. 

The above scenario is generic in SS codes \cite{barbier2017approximate,biyik2017generalized} (and in high-dimensional inference more generically \cite{barbier2019optimal}), but it is also possible that no hard region is present at all (i.e., $R_\text{IT}=R_\text{GVAMP}$ and a single fixed point of the stationary equations exists for all rates). E.g., this happens at low SNR and/or low section size $B$ for the AWGN channel. See \cite{barbier2017approximate} for the same phenomenology and plots for visualization. 

\begin{table}[t]
    \renewcommand{\arraystretch}{1.2}
    \setlength\tabcolsep{7pt}
	\centering{ 
	\vspace*{-5pt}
    \begin{threeparttable}
	\begin{tabular}{|c|c|c|c|c|}
	\hline
    \textbf{BEC} & $R^g_{\text{GVAMP}}$ & $R^g_{\text{IT}}$ & $R^r_{\text{GVAMP}}$ & $R^r_{\text{IT}}$ \\
    \hline
	B=2 & $0.428$ & $0.511$ & $0.481$ & $0.553$     \\
    \hline
    B=4 & $0.546$ & $0.662$ & $0.603$ & $0.713$     \\
    \hline   
    B=8 & $0.607$ & $0.748$ & $0.657$ & $0.783$     \\
	\hline

	\textbf{BSC} & $R^g_{\text{GVAMP}}$ & $R^g_{\text{IT}}$ & $R^r_{\text{GVAMP}}$ & $R^r_{\text{IT}}$ \\
	\hline
	B=2 & $0.426$ & $0.513$ & $0.468$ & $0.552$    \\
    \hline
    B=4 & $0.545$ & $0.663$ & $0.602$ & $0.715$    \\
    \hline
    B=8 & $0.612$ & $0.743$ & $0.662$ & $0.794$    \\
    \hline

	\textbf{ZC}& $R^g_{\text{GVAMP}}$ & $R^g_{\text{IT}}$ & $R^r_{\text{GVAMP}}$ & $R^r_{\text{IT}}$ \\
	\hline
    B=2 & $0.396$ & $0.475$ & $0.432$ & $0.515$    \\
    \hline
    B=4 & $0.507$ & $0.618$ & $0.556$ & $0.664$    \\
    \hline
    B=8 & $0.565$ & $0.693$ & $0.615$ & $0.742$    \\
    \hline
	\end{tabular}
	\end{threeparttable}
	}
	\vspace{10pt}
	\caption{{GVAMP threshold  $ R_{\text{GVAMP}}  $ and   information theoretic threshold  $ R_{\text{IT}}$. The error probability $\epsilon$ for BEC, BSC and ZC is $0.1$, $0.01$ and $0.05$ respectively. Superscript $g$ and $r$ signify Gaussian matrix and row-orthogonal matrix respectively. Subscripts IT and GVAMP index the information-theoretic  and algorithm thresholds, respectively.}}\label{table:gampForSS}
	\vspace{-20pt}
\end{table}

\section{Vanishing error floor property for $B<\infty$}\label{sec:4} The MSE floor $E_{f}$ is the MSE attained from the informative initialization \cite{barbier2016threshold,biyik2017generalized}: $E_{f}:=1-B{q}_{{\rm in},x}^\infty= E(\hat{q}_{{\rm in},x}^\infty)$. It matches GVAMP performance and the MMSE in the easy phase and the MMSE only in the hard phase, while it has no concrete meaning in the impossible phase. In \cite{biyik2017generalized} it is shown that as $L \to +\infty$ the error floor vanishes for any $R$ and $B$ for a wide class of binary inputs channels, but only for i.i.d. Gaussian coding matrices. We heuristically show that \emph{the vanishing error-floor property universally holds for rotationally invariant coding matrices with compactly supported spectra}. We focus on the BEC channel but the approach can be generalized to other binary input channels.

Our strategy is to \emph{assume} both $i)$ the existence of a solution to the stationary equations such that $E_f=0$, namely, such that ${q}_{{\rm in},x}^\infty=B^{-1}$, and look for a self-consistent set of parameters values for the remaining stationary equations, and $ii)$ this potential solution is such that the corresponding free entropy ($\Phi_{\rm RS}$ evaluated in it) is the largest when a second solution exists, and this for all $R<R_{\text{IT}}$. Point $ii)$ as well as the fact that only two solutions may co-exist have been thoroughly numerically verified in the present setting and previous ones \cite{biyik2017generalized}. We now simply denote $\mathbf{q}_{\rm in}$ by $(q_{x}, q_{z},\hat{q}_{x},\hat{q}_{z})$. So we reverse engineer the solution starting from $q_x=B^{-1}$. Recall \eqref{MSE}. From $E(\hat q_x)=1-Bq_x=0$,  $q_x=B^{-1}$ requires $\hat q_x=+\infty$ (this can be seen from the $q_x$-stationary equation too). When setting $(q_x=B^{-1},\hat q_x=+\infty)$ in the $\hat q_x$-stationary equation we further deduce that $q_z=+\infty$. This is easily seen in the MP case: using $\mathcal{F}_{\text{MP}}(x,y)=-\frac{\alpha}2 xy$ then $2 \partial q_{x} \mathcal{F}(1-B q_{x}, q_{z})= B\alpha q_z$. Thus the $\hat q_x$-stationary equation becomes $+\infty=B\alpha q_z$ which implies $q_z=+\infty$. For general spectral law we use that \cite{maillard2020construction} $\mathcal{F}(x,y)= \mathcal{F}_{\text{MP}}(x,y) + O(x^2)$. And because $1-B q_{x}\to 0$ around the desired solution $q_x=B^{-1}$ the same argument applies: for any spectrum $(q_x=B^{-1},\hat q_x=+\infty)$ implies $q_z=+\infty$. We finally need to fix the $\hat q_z$ using the $\hat q_z$-stationary equation. Using the same approach, $\hat q_z=1+2\alpha^{-1}\partial q_{z} \mathcal{F}_{\text{MP}}(1-B q_{x}, q_{z})=1-(1-B q_{x})=Bq_x$ at the desired solution. So $(q_x=B^{-1},\hat q_x=+\infty, q_z=+\infty,\hat q_z=1)$ is a self-consistent solution \emph{if and only if} it also verifies the last equation we did not exploit, namely the $q_z$-stationary equation. 

Let $\mathcal{D} z$ be the standard Gaussian measure. For the BEC channel the right-hand side of the $q_z$-stationary equation is 
\begin{equation*}
\textstyle{(1-\epsilon)(2 \pi \sqrt{1-\hat{q}_z})^{-1} \int \mathcal{D} z \exp(-\frac{1}{2} \hat{q}_z z^2)(\int_{\sqrt{\hat{q}_z} z}^\infty\mathcal{D} x)^{-1}.}
\end{equation*}
As $\hat{q}_z \to 1$ it diverges, meaning $q_z \to +\infty$. Consequently, $\mathbf{q}_{\rm in}=(q_x=B^{-1},\hat q_x=+\infty, q_z=+\infty,\hat q_z=1)$ is solution of the stationary equations for the BEC channel and any $\rho$, and thus $E_f=0$. The same argument can be extended to other binary input channels (BSC, ZC, etc.); we confirm this numerically in Fig. \ref{fig}. To show it analytically, the concrete expressions of the (right-hand side of the) $q_z$-stationary equation for other channels are found in table $\uppercase\expandafter{\romannumeral1}$ of \cite{biyik2017generalized} (with the variable substitution $1-\hat{q}_z\to E$). Instead, for  the AWGNC with signal-to-noise ratio $\gamma$ the $q_z$-stationary equation is $\gamma/(1+ \gamma(1- \hat{q}_z))$ which does \emph{not} diverge when $\hat q_x \to 1$. Thus $E_f>0$; however  $\lim_{B\to +\infty} E_f=0$, see \cite{barbier2017approximate}.

\section{Achieving the capacity as $B\to+\infty$}
\label{sec:larg_B}

We now show that as $B\to+\infty$ (after $L\to+\infty$) the threshold $R_{\text{IT}}$ tends to the Shannon capacity $C$ for binary input channels, whenever a simple ``spectral criterion'' is verified:


\begin{theorem}
Consider SS codes for any memoryless channel. Let the coding matrix $\mathbf{A}$ be drawn from a rotational invariant ensemble, and whose empirical spectral measure converges to a well defined density with finite support as $L\to \infty$. The code is capacity achieving in the sense that $ \lim_{B \to \infty} R_{\text{IT}}=C$ if and only if the asymptotic p.d.f. $\rho_{\text{supp}}$ of the non-zero eigenvalues of $B^{-1} \mathbf{A}^\intercal \mathbf{A}$ verifies $\rho_{\text{supp}} \to \delta_1$ in law when $B\to\infty$, $\alpha \to 0$.
\end{theorem}
According to this principle both the Gaussian and row-orthogonal ensembles are capacity-achieving as $L\to+\infty$ followed by $B\to +\infty$. For the row-orthogonal case this spectral criterion is even satisfied for finite $B$, which may explain its improved performance at finite section size. Note that for $R$ to remain finite in this limit then necessarily $\alpha =\Theta(\ln B /B)\to 0$. The threshold $R_{\text{IT}}$ for finite section size $B$ shown in Table I converges when $B$ increases to the predicted limit. Result 1 is based on the analysis of the rescaled potential $\tilde \Phi_{\rm RS}:= \Phi_{\rm RS}/{\ln B}$. One needs also to define rescaled parameters $r_x:= B q_x$ and $\hat r_x:= \hat q_x/\ln B$ as in \cite{barbier2017approximate}. All the rescaled quantities have non-trivial limits as $B\to+\infty$. We propose an heuristic, numerically verified, argument showing that \emph{as $B\to+\infty$ the potential $\tilde{\Phi}_{\rm RS}$ possesses only two maxima, one verifying $r_x= 1$ and another $r_x= 0$} (see \cite{barbier2017approximate,hou2022sparse} for related arguments). The same holds with Gaussian coding matrices \cite{barbier2017approximate,barbier2016proof,barbier2016threshold}. Indeed, the only term dependent on $\hat r_x$ in $\tilde \Phi_{\rm RS}$ is  
\begin{equation}
 \textstyle{\tilde I_0(r_x,\hat r_x):= I_0/\ln B \to \max(1,  \hat{r}_x/2)- r_x \hat{r}_x/ 2 }
\end{equation}
as $B\to+\infty$; this was computed in \cite{barbier2017approximate}. Considering its $\hat r_x$-derivative to obtain the $r_x$-stationary equation, and given that $r_x\in[0,1]$, it is clear that for $r_x$ to possibly change its value (i.e., existence of two solutions) the ``effective signal-to-noise'' (SNR) $\hat r_x$ must transition at $\hat{r}_x=2$ whatever is the solution of the $\tilde \Phi$-stationary equations for the remaining  parameters. So two scenarios are possible:

\noindent\textbf{High error case}: The effective SNR $\hat r_x$ solution to the $\tilde \Phi_{\rm RS}$-stationary equations is low enough so that $\max(1, \frac{\hat{r}_x}{2}) = 1$. Then the $r_x$-stationary equation obtained by setting $\partial_{\hat r_x}\tilde I_0=0$ enforces $r_x=0$ meaning no decoding at all (recall the link between overlap and MSE).

\noindent\textbf{No error case}: This time the solution $\hat r_x$ is large enough so that $\max(1, \frac{\hat{r}_x}{2}) = \frac{\hat{r}_x}{2}$. Then $\partial_{\hat r_x}\tilde I_0=0$ yields the second solution $r_x=1$, i.e., perfect decoding.

We argued that only two solutions exist and can now derive Result 1. From the definition of $R_{\text{IT}}$, we look for a rate such that $\tilde \Phi_{\rm RS}(r_x=0)=\tilde \Phi_{\rm RS}(r_x=1)$ (the other parameters being understood to be set at their respective solutions). 

In the \textbf{no error case} $r_x=1$ it is direct to see that $\mathcal{F}(x,y)$ is independent of $\rho$ and thus $\mathcal{F}=\mathcal{F}_{\text{MP}}(x,y)$. As this is the only $\rho$-dependent part of the potential $\tilde \Phi_{\rm RS}(r_x=1)=\tilde \Phi^{\text{MP}}_{\rm RS}(r_x=1)$, the rescaled potential when considering the MP law $\rho$. As explained above, we also have $\tilde{I}_{0}(r_x=1,\hat r_x) = 0$. We have seen in Sec.~\ref{sec:4} that perfect decoding implied the solution $\hat q_z=1$ and $q_z=+\infty$ for any $B$ (and thus in the limit). All-in-all it yields  $\tilde{\Phi}_{\rm RS}(r_x=1)=\tilde{\Phi}_{\rm RS}^{\text{MP}}=\frac{1}{R}\mathbb{E}_{z\sim\mathcal{N}(0,1)}\int dyP_{\text{out}}( y\mid z) \log_2 P_{\text{out}}( y\mid z )$ for this non error solution, and for any $\rho$.

Now the \textbf{high error case} $r_x=0$. By [Lemma 1, \cite{hou2022sparse}] the \emph{R-transform} $\mathcal{R}(x)$ associated to a generic $\rho$ \cite{tulino2004random} is upper bounded, when $B\to+\infty$, $\alpha\to 0$, by the one of the MP law. Then using the equivalent expression $\mathcal{F}(x,y) = \frac12\inf_{\Lambda_y > 0}\{ \int_0^{-\frac{x}{\Lambda_y}} \mathcal{R}(t) d t + \alpha \Lambda_y y -  \alpha \ln \Lambda_y - \alpha \ln y -  \alpha\}$ we can show $ \tilde{I}_{\text{int}} \geq \tilde{I}^{\text{MP}}_{\text{int}}$ where $ \tilde{I}_{\text{int}}=\lim_{B}  I_{\text{int}} /\ln B$. Because $\tilde{I}_{\text{int}}$ is the only spectrum-dependent term of the potential we automatically deduce $\tilde{\Phi}_{\rm RS} \geq \tilde{\Phi}_{\rm RS}^{\text{MP}}$ when evaluated at the same solution (the high error one in that particular case). Equality holds if and only if $\rho  \to \alpha \delta_1 + (1-\alpha) \delta_0$ in law as $\alpha \to 0$.
In the i.i.d. Gaussian/MP ensemble $ \mathcal{F}(x,y)=-\frac{\alpha}{2}xy $ which implies, when $r_x=0$,  $\tilde I^{\text{MP}}_{\text{int}} = 0$ independently of $q_z$. Because we also have $\tilde I_0=1$ for the high error solution, the lower bound on the replica potential reads $ \tilde{\Phi}_{\rm RS}^{\text{MP}}(r_x\!=\!0)\!=\!1\!+\! \frac{1}{R} \int d y\left(\mathbb{E}_{z} P_{\text {out }}(y \mid z)\right) \log_{2}\mathbb{E}_{z} P_{\text {out }}(y \mid z)$. 
From this ``high error lower bound'' $ \tilde{\Phi}_{\rm RS} =\epsilon_{\rho} +\tilde{\Phi}_{\rm RS}^{\text{MP}}(r_x=0)$ where $\epsilon_{\rho}\geq 0$, with equality if and only if $\rho \to \alpha \delta_1 + (1- \alpha) \delta_0$ as $\alpha \to 0$. 

$R_{\text{IT}}$ is obtained by solving $\tilde{\Phi}_{\rm RS}(r_x=0)= \tilde{\Phi}_{\rm RS}(r_x=1)$. This yields $R_{\text{IT}}=\frac{1}{1+\epsilon_{\rho}} \big[ \mathbb{E}_{z} \int d y P_{\text{out}}(y|z) \ln _{2} P_{\text{out}}(y|z)- \int d y\left(\mathbb{E}_{z} P_{\text{out}}(y|z)\right) \ln _{2}\left(\mathbb{E}_{z} P_{\text{out}}(y|z)\right)   \big]=C/(1+\epsilon_{\rho})$.  The coding scheme is thus capacity-achieving if and only if $\epsilon_{\rho} = 0$, i.e., $ \rho_{\text{supp}} \to \delta_{1}$ in law  when $\alpha \to 0 $. This ends the argument.


\section*{Acknowledgments}
 J.B and H.T.Q thank  Nicolas Macris and  Liang Shansuo   for helpful discussions.
\ifCLASSOPTIONcaptionsoff
\fi
%


%
%
\end{document}

%% file: settings_custom.tex
\DeclareUnicodeCharacter{00A0}{~}

\def \({\left(}
\def \){\right)}
\def \[{\left[}
\def \]{\right]}
\newcommand{\tbf}[1]{{\mathbf{#1}}}

\newcommand{\defeq}{\vcentcolon=}

\newcommand{\br}{{\mathbf {r}}}

\newcommand{\bw}{{\mathbf {w}}}

\newcommand{\bA}{{\mathbf {A}}}

\newcommand{\bx}{{\mathbf {x}}}

\newcommand{\by}{{\mathbf {y}}}
\newcommand{\bz}{{\mathbf {z}}}

\newcommand{\bs}{{\mathbf {s}}}
\newcommand{\bS}{{\mathbf {S}}}

\newcommand{\bxi}{{\boldsymbol{\xi}}}

\newcommand{\be}{\begin{equation}}
\newcommand{\ee}{\end{equation}}
\newcommand{\bea}{\begin{eqnarray}}
\newcommand{\eea}{\end{eqnarray}}

\newcommand{\mat}[1]{\ensuremath{\begin{bmatrix}#1\end{bmatrix}}}
\newcommand{\smallmat}[1]{\ensuremath{
        \left[\begin{smallmatrix}#1\end{smallmatrix}\right]}}

\newtheorem{theorem}{Result}

